\documentclass{article}
\usepackage{color}
\usepackage{graphicx}
\usepackage{amsmath, amsthm}
\usepackage{mathtools}
\usepackage{natbib}
\usepackage{url}
\RequirePackage[colorlinks,citecolor=blue,urlcolor=blue]{hyperref}
\usepackage{xcolor}
\usepackage{chngcntr}
\usepackage{subfig}
\usepackage{hyperref}
\usepackage{algorithm}
\usepackage[noend]{algpseudocode}
\usepackage{setspace}
\newtheorem{theorem}{Theorem}

\usepackage{xfrac}
\definecolor{forest}{rgb}{0.133,0.545,0.133}
\usepackage{multirow}
\usepackage{amsfonts}
\newtheorem{lemma}{Lemma}

\usepackage{enumitem}
\usepackage{caption}

\evensidemargin=\oddsidemargin
\usepackage{etoolbox}

\newif\ifabbreviation
\pretocmd{\thebibliography}{\abbreviationfalse}{}{}
\AtBeginDocument{\abbreviationtrue}

\begin{document}
	\newcommand{\bb}{\boldsymbol{\beta}}

	\title{Group Sequential Design with Posterior and Posterior Predictive Probabilities}


	\author{Luke Hagar\footnote{Luke Hagar is the corresponding author and may be contacted at \url{lmhagar@uwaterloo.ca}.} \hspace{35pt} Shirin Golchi$^+$ \hspace{25pt} Marina B. Klein$^{\dagger}$ \bigskip \\ 
    $^*$\textit{Clinical Trials Capability, The University of Queensland} \\
$^+$\textit{Department of Epidemiology, Biostatistics \& Occupational Health, McGill University} \\
$^{\dagger}$\textit{McGill University Health Centre, McGill University}}

	\date{}

	\maketitle

	\begin{abstract}

Group sequential designs drive innovation in clinical, industrial, and corporate settings. Early stopping for failure in sequential designs conserves experimental resources, whereas early stopping for success accelerates access to improved interventions. Bayesian decision procedures provide a formal and intuitive framework for early stopping using posterior and posterior predictive probabilities. Design parameters including decision thresholds and sample sizes are chosen to control the error probabilities associated with the sequential decision process. These choices are routinely made based on estimating the sampling distribution of posterior summaries via intensive Monte Carlo simulations for each sample size and design scenario considered. In this paper, we propose an efficient method to calibrate decision thresholds to pre-specified alpha- and beta-spending functions and determine minimum sample sizes for Bayesian group sequential designs. We prove theoretical results that enable posterior and posterior predictive probabilities to be modeled as a function of the sample size. Using these functions, we assess error probabilities at a range of sample sizes given simulations conducted at only two sample sizes. The effectiveness of our methodology is highlighted using several substantive examples.

		\bigskip

		\noindent \textbf{Keywords:}
		Bayesian sample size determination; clinical trials; dynamic borrowing; experimental design; quality control
	\end{abstract}

	\maketitle

	\baselineskip=19.5pt



\section{Introduction}\label{sec:intro}

 Scientific experiments drive progress across a wide range of disciplines, including medicine, manufacturing, public transportation, and consumer marketing.
 While exploring new interventions to assess their suitability is crucial to foster innovation, ethical and economic concerns arise when accruing knowledge is not exploited to offer people the best available intervention. Superior interventions should be implemented and inferior ones should be discontinued as quickly as possible. Sequential designs \citep{wald2004sequential} can mitigate the ethical concerns and financial costs of experimentation. Group sequential designs (GSDs) divide experiments into stages, analyzing data after each stage to decide whether to continue based on predefined discontinuation rules \citep{bross1952sequential,jennison1999group}. 
Early stopping for success is based on evidence from the data that a new intervention is beneficial, whereas early stopping for failure is based on evidence of ineffectiveness. Competing null and alternative hypotheses -- $H_0$ and $H_1$ -- respectively characterize settings where a new intervention is ineffective and beneficial. 

To ensure GSDs reliably inform decision making, the error probabilities associated with their decision procedures must be controlled. These error probabilities are the probabilities of making incorrect decisions across all analyses, such as stopping for success when $H_0$ is true or not stopping for success under $H_1$. The error probabilities of sequential designs are typically controlled by selecting suitable sample sizes and decision thresholds for the repeated analyses. Popular methods for selecting decision thresholds adjust for the inflated type I error risk linked to early stopping for success in GSDs \citep{pocock1977group,o1979multiple, demets1994interim}. \color{black} Decision thresholds related to early stopping for failure have historically been chosen using stochastic curtailment procedures \citep{halperin1982aid,spiegelhalter1986monitoring,lachin2005review}. These procedures advocate for stopping if there is a low probability of achieving the experiment's objective in its remaining stages given the available data and assumptions about the future data. The aforementioned methods were developed with a primary focus in frequentist design of experiments. \color{black}

Bayesian methods provide a formal and intuitive framework for early stopping in sequential designs based on posterior summaries. For example, the experiment can be stopped for success at any analysis if the posterior probability that $H_1$ is true exceeds the corresponding decision threshold. Posterior predictive probabilities \citep{rubin1984bayesianly,gelman1996posterior,berry2010bayesian,saville2014utility} quantify the probability that the posterior probability that $H_1$ is true at a future analysis exceeds its decision threshold. The experiment can be stopped for success or failure depending on whether this posterior predictive probability is sufficiently large or small. Posterior predictive probabilities serve as a Bayesian analog to stochastic curtailment with fewer explicit assumptions about the future data, which are generated based on the current posterior. 

Even when Bayesian posterior summaries inform decision making, sample sizes and decision thresholds are often chosen to control the frequentist error probabilities of sequential designs. Clinical regulators require strict control of error probabilities \citep{fda2019adaptive}, but the frequentist error probabilities of Bayesian designs are of much broader interest (see e.g., \citet{jenkins2011power, deng2024metric}). \color{black} Sample sizes and decision thresholds that approximately attain criteria for the frequentist error rates of Bayesian GSDs could be chosen using numeric quadrature \citep{armitage1969repeated, jennison1999group}. However, such methods rely on asymptotic normal approximations that are unsuitable when using informative prior distributions, a feature increasingly incorporated into Bayesian clinical trials and GSDs more generally  \citep{fda2026bayesian}.  \color{black} 

\citet{wang2002simulation} proposed a flexible framework to determine suitable sample sizes and decision thresholds that uses Monte Carlo simulation to estimate error probabilities of Bayesian designs. \color{black}This computational approach accounts for informative priors and small-sample deviations from asymptotic results by \color{black}estimating sampling distributions of posterior summaries through simulating many iterations of an experiment according to a particular data generation process. \citet{gubbiotti2011bayesian} defined two methodologies for specifying the data generation process; the conditional approach uses fixed values for the data generating parameters and the predictive approach accommodates uncertainty in these values. With either approach, the repeated estimation of joint sampling distributions across the stages of a sequential design requires substantial computing resources.  A general and efficient method for GSD with posterior and posterior predictive probabilities would make these designs more accessible to practitioners.

Various strategies have recently been proposed to reduce the computational overhead required to estimate sampling distributions of posterior summaries. \citet{golchi2022estimating} and \citet{golchi2024estimating} proposed flexible modeling approaches to estimate sampling distributions of univariate summaries. \citet{hagar2025scalable} developed a method to estimate the sampling distribution of posterior probabilities throughout the sample size space using estimates of the sampling distribution at only two sample sizes. Because these approaches do not consider the joint sampling distribution of posterior summaries across multiple analyses, they are not suitable for sequential designs. In this work, we build upon the method from \citet{hagar2025scalable} to accommodate both GSDs and the use of posterior predictive probabilities. These useful extensions are predicated on a series of theoretical results that are original to this paper. While our framework is theoretically intricate, its straightforward implementation promotes an economical and broadly useful approach to simulation-based design for Bayesian group sequential experiments. 

The remainder of this article is structured as follows. In Section \ref{sec:prelim}, we introduce preliminary concepts required to describe our methods. In Section \ref{sec:methods}, we construct proxies to the joint sampling distribution of posterior and posterior predictive probabilities across multiple analyses  and prove novel theoretical results about these proxies. We adapt these theoretical results to develop a procedure to select sample sizes and decision thresholds for GSDs  that requires estimation of a true joint sampling distribution of posterior and posterior predictive probabilities at only two sample sizes in Sections \ref{sec:stop} and \ref{sec:power}. In Section \ref{sec:ex}, we showcase the strong performance of our methodology with \color{black} an example clinical trial that dynamically borrows historical information using an informative prior. \color{black} We  discuss extensions to this work in Section \ref{sec:disc}. \color{black} Additional examples provided in the online supplement illustrate the broad applicability of our methodology. \color{black}

  \section{Preliminaries}\label{sec:prelim}

This paper focuses on Bayesian GSDs with at most two interventions in which predefined stopping rules leverage posterior and posterior predictive probabilities about the target of inference. The statistical model for the experiment is defined via parameters $\boldsymbol{\theta} \in \boldsymbol{\Theta}$. The target of inference is a function of these parameters: $\delta(\boldsymbol{\theta}) \in \mathbb{R}$. The interval hypotheses that inform decision making are $H_0: \delta(\boldsymbol{\theta}) \notin (\delta_L, \delta_U)$ vs.\ $H_1: \delta(\boldsymbol{\theta}) \in (\delta_L, \delta_U)$,
where $-\infty \le \delta_L < \delta_U \le \infty$. This general notation for the interval endpoints accommodates comparisons based on superiority, noninferiority, and practical equivalence. The posterior distribution of $\delta(\boldsymbol{\theta})$ synthesizes information from the prior distribution for $\boldsymbol{\theta}$ and the available data.  

Sequential experiments have $T$ potential analyses indexed by $t \in \{1, \dots, T\}$. For theoretical development, we first consider a hypothetical GSD that cannot be terminated before analysis $T$. Posterior and posterior predictive probabilities are observed at analyses $t < T$ in such a design, but they do not inform decision making. In Section \ref{sec:stop}, we use our results for GSDs without early stopping to accommodate designs that can stop for both success and failure at analyses $t < T$. At the $t^{\text{th}}$ analysis for a design without early stopping, the cumulative $n_t$ observations comprise the data, $\mathcal{D}_{n_t} = \{{\bf{Y}}_{n_t \times 1}, {\bf{X}}_{n_t \times w} \}$, consisting of outcomes ${\bf{Y}}_{n_t \times 1}$ and $w$ additional covariates ${\bf{X}}_{n_t \times w}$ for each observation. All observations accrued in previous stages are retained in $\mathcal{D}_{n_t}$ for subsequent analyses. 



 For a design that precludes early stopping,  we index the joint sampling distribution of posterior summaries using the sample size for the first analysis. Specifically, we index by a sample size $n$ such that $\{n_1, n_2, \dots, n_T\} = n \times \{c_1, c_2, \dots, c_T\}$ for some constants $c_1=1$ and $\{c_t\}_{t=2}^T > 1$.  \color{black} For a design with two interventions, we suppose allocation probabilities $\{A_t\}_{t=1}^T \in [0,1]^T$ are pre-specified such that $100\times A_t\%$ of observations are randomized to a particular intervention during stage $t$. \color{black} The data $\mathcal{D}_n = \{\mathcal{D}_{n_t} \}_{t=1}^T$ across all analyses define a vector of $T$ posterior probabilities about the hypothesis $H_1$:
  \begin{equation}\label{eq:pp}
           \boldsymbol{\tau}(\mathcal{D}_{n})  = 
           \begin{bmatrix}
           \tau_1(\mathcal{D}_{n}) \\
           \vdots \\
           \tau_T(\mathcal{D}_{n}) 
         \end{bmatrix} = \begin{bmatrix}
           \Pr(H_{1}~|~\mathcal{D}_{n_1}) \\
           \vdots \\
           \Pr(H_{1}~|~\mathcal{D}_{n_T}) 
         \end{bmatrix}.
\end{equation} 
We conclude success at analysis $T$ if $\tau_T(\mathcal{D}_{n}) \ge \gamma_T$ for some success threshold $\gamma_T \in [0,1]$.  

We next consider posterior predictive probabilities. The posterior predictive distribution characterizes the distribution of future data according to the current posterior distribution \citep{rubin1984bayesianly, gelman1996posterior}. The posterior predictive distribution $p_{\hspace*{0.75pt}\text{P}}(y~|~\mathcal{D}_{n_t})$ at analysis $t$ is defined as
  \begin{equation}\label{eq:pred.dist}   
           p_{\hspace*{0.75pt}\text{P}}(y~|~\mathcal{D}_{n_t})= \int_{\boldsymbol{\Theta}} p(y~|~\boldsymbol{\theta})p\left(\boldsymbol{\theta}~|~\mathcal{D}_{n_{t}}\right)d\boldsymbol{\theta},
\end{equation} 
where $p(y~|~\boldsymbol{\theta})$ is the assumed model for the outcome and  $p\left(\boldsymbol{\theta}~|~\mathcal{D}_{n_{t}}\right)$ is the current posterior distribution.
Posterior predictive probabilities generally represent probability statements about future data or parameters that are inferred from a posterior that incorporates these future data. Following common practice \citep{berry2010bayesian,saville2014utility}, we specifically consider the posterior predictive probability for an analysis $t < T$ as the probability that $\tau_T(\mathcal{D}_{n})$ will be at least $\gamma_T$ given the current data $\mathcal{D}_{n_t}$. This probability is considered when the data generation process for the remaining $n_T - n_t$ observations is the posterior predictive distribution in (\ref{eq:pred.dist}):
  \begin{equation}\label{eq:ppp.def}   
           \Pr\{\Pr(H_{1}~|~\mathcal{D}_{n_T}) \ge \gamma_{T}~|~\mathcal{D}_{n_{t}}\} = \int_{\boldsymbol{\Theta}} \Pr\{\Pr(H_{1}~|~\mathcal{D}_{n_T}) \ge \gamma_{T}~|~\boldsymbol{\theta}\}p\left(\boldsymbol{\theta}~|~\mathcal{D}_{n_{t}}\right)d\boldsymbol{\theta}.
\end{equation}

The posterior predictive distribution can be obtained analytically for simple models with conjugate priors \citep{saville2014utility}, but the \emph{intensive} simulation-based procedure that follows is generally applicable. First, a sample value $\boldsymbol{\theta}^{_{(m)}}$ is drawn from the posterior distribution $p\left(\boldsymbol{\theta}~|~\mathcal{D}_{n_{t}}\right)$. Second, $n_T - n_t$ observations are generated from the assumed model $p(y~|~\boldsymbol{\theta}^{_{(m)}})$ and combined with $\mathcal{D}_{n_{t}}$ to create $\mathcal{D}_{n_{T}}^{_{(m)}}$. Third, the posterior distribution $p\left(\boldsymbol{\theta}~|~\mathcal{D}_{n_{T}}^{_{(m)}}\right)$ is approximated to compute the posterior probability $\Pr(H_{1}~|~\mathcal{D}_{n_T}^{_{(m)}})$. These three steps are repeated $M$ times. The posterior predictive probability in (\ref{eq:ppp.def}) is estimated as $M^{-1}\sum_{m = 1}^M\mathbb{I}\{\Pr(H_{1}~|~\mathcal{D}_{n_T}^{_{(m)}}) \ge \gamma_T\}$.

The data, indexed by the sample size $n$ for the first analysis, define a vector of $T-1$ posterior predictive probabilities:
  \begin{equation}\label{eq:ppp}
           \boldsymbol{\tau}_{\text{P}}(\mathcal{D}_{n})
    = \begin{bmatrix}
           \tau_{\text{P}, 1}(\mathcal{D}_n) \\
           \vdots \\
           \tau_{\text{P}, T-1}(\mathcal{D}_n)
         \end{bmatrix} = \begin{bmatrix}
           \Pr\{\Pr(H_{1}~|~\mathcal{D}_{n_T}) \ge \gamma_{T}~|~\mathcal{D}_{n_{1}}\} \\
           \vdots \\
           \Pr\{\Pr(H_{1}~|~\mathcal{D}_{n_T}) \ge \gamma_{T}~|~\mathcal{D}_{n_{T-1}}\}
         \end{bmatrix}.
\end{equation} 

These posterior predictive probabilities will later be used to inform interim decisions in GSDs with early stopping. As detailed in Section \ref{sec:stop}, we can assess error probabilities in sequential designs \emph{with} early stopping under various scenarios using the joint sampling distribution of the posterior probabilities in (\ref{eq:pp}) and the posterior predictive probabilities in (\ref{eq:ppp}). We jointly refer to these probabilities as
  \begin{equation*}\label{eq:both}
           \boldsymbol{\tau}_{\hspace*{-1.25pt}*}(\mathcal{D}_{n})
    = \begin{bmatrix}
           \boldsymbol{\tau}(\mathcal{D}_{n}) \\
           \boldsymbol{\tau}_{\text{P}}(\mathcal{D}_{n})
         \end{bmatrix}.
\end{equation*} 

To estimate the sampling distribution of $\boldsymbol{\tau}_{\hspace*{-1.25pt}*}(\mathcal{D}_{n})$ via simulation, we define various data generation processes for $\mathcal{D}_{n}$. For each Monte Carlo iteration, data are generated according to a fixed parameter value $\boldsymbol{\theta}$.  The probability model $\Psi$ characterizes how $\boldsymbol{\theta}$ values are drawn in iteration $r = 1, \dots, R$. This probability model $\Psi$ characterizes a scenario under which we later assess the error probabilities of GSDs; it could be viewed as a \emph{design} prior \citep{gubbiotti2011bayesian} that differs from the \emph{analysis} prior $p(\boldsymbol{\theta})$. This notation accommodates the conditional and predictive approaches, where $\Psi$ is a degenerate probability model under the conditional approach. For each iteration $r$, data $\mathcal{D}_{n, r}$ across all $T$ stages are generated given $\boldsymbol{\theta}_{r} \sim \Psi$, and $\boldsymbol{\tau}_{\hspace*{-1.25pt}*}(\mathcal{D}_{n, r})$ is computed. Across $R$ iterations, the collection of obtained $\{\boldsymbol{\tau}_{\hspace*{-1.25pt}*}(\mathcal{D}_{n, r})\}_{r = 1}^R$ values estimates the joint sampling distribution of $\boldsymbol{\tau}_{\hspace*{-1.25pt}*}(\mathcal{D}_{n})$.

In Section \ref{sec:stop}, we describe how the sample sizes $\{n_t\}_{t=1}^T$ and decision thresholds impact the error probabilities of GSDs with early stopping. For every value of $n = n_1$ considered, we must obtain a collection of $\{\boldsymbol{\tau}_{\hspace*{-1.25pt}*}(\mathcal{D}_{n, r})\}_{r = 1}^R$ values via simulation to estimate error probabilities, while tuning the decision thresholds, to find a suitable design. The process to obtain the $\{\boldsymbol{\tau}_{\hspace*{-1.25pt}*}(\mathcal{D}_{n, r})\}_{r = 1}^R$ values is computationally intensive. However, we can reduce the computational burden by using previously estimated sampling distributions of $\boldsymbol{\tau}_{\hspace*{-1.25pt}*}(\mathcal{D}_{n})$ to estimate error probabilities at new $n$ values without conducting additional simulations. We can use this process to efficiently design Bayesian sequential experiments. We propose such a design method in this paper and begin its development in Section \ref{sec:methods}. 

  \section{Proxies to the Joint Sampling Distribution}\label{sec:methods}

\subsection{Proxies to Posterior Probabilities}\label{sec:methods.proxy1}

To motivate our design procedure proposed in Section \ref{sec:power}, we create a proxy to the joint sampling distribution of $\boldsymbol{\tau}_{\hspace*{-1.25pt}*}(\mathcal{D}_{n})$ for a hypothetical design without early stopping. These proxies are needed for the theory that substantiates our proposed methodology. However, our methods do not directly use these proxies and instead estimate the true sampling distribution of $\boldsymbol{\tau}_{\hspace*{-1.25pt}*}(\mathcal{D}_{n})$ by simulating samples $\{\mathcal{D}_{n, r}\}_{r=1}^R$ and approximating posterior summaries as described in Section \ref{sec:prelim}. We create a proxy for the joint sampling distribution of posterior probabilities $\boldsymbol{\tau}(\mathcal{D}_{n})$ in this subsection and augment this proxy to accommodate posterior predictive probabilities in Section \ref{sec:methods.proxy2}. 

For reasons described shortly, our proxies leverage the regularity conditions for the asymptotic normality of the maximum likelihood estimator (MLE), detailed in Theorem 5.39 of \citet{vaart1998bvm}. To design group sequential experiments, we consider the joint sampling distribution of the MLE $\hat{\boldsymbol{\delta}}^{_{(n)}}_r = \{\hat{\delta}^{_{(n)}}_{t,r}\}_{t=1}^T$ across all analyses, where the first subscript in $\hat{\delta}^{_{(n)}}_{t,r}$ indexes the analysis and the second denotes the Monte Carlo iteration. We index all components of the MLE $\hat{\boldsymbol{\delta}}^{_{(n)}}_r$ by the sample size $n = n_1$ for the first analysis, but note that the MLE $\hat{\delta}^{_{(n)}}_{t,r}$ is in fact based on a total of $n_t = c_tn$ observations. These constants $\{c_t\}_{t=1}^T$ and the constants $\{A_t\}_{t=1}^T$ for the pre-specified allocation probabilities are omitted from our notation for the joint MLE for simplicity. Under these conditions, Lemma \ref{lem0} establishes the approximate joint sampling distribution of the MLE,  $\hat{\boldsymbol{\delta}}^{_{(n)}} ~|~ \boldsymbol{\theta} = \boldsymbol{\theta}_{r}$, given some true parameter values $\boldsymbol{\theta}_{r} \sim \Psi$ in iteration $r$. 

 \begin{lemma}\label{lem0}
 For any $\boldsymbol{\theta}_{r} \sim \Psi$, let the model $p(y~|~\boldsymbol{\theta}_r)$ satisfy the conditions for the asymptotic normality of the MLE for a design without early stopping. \color{black} We suppose the allocation probabilities $\{A_t\}_{t=1}^T \in [0,1]^T$ are pre-specified \color{black} and the cumulative sample sizes for the analyses are such that $\{n_t\}_{t=1}^T = n \times \{1, c_2, \dots, c_T\}$ for constants $\{c_t\}_{t=2}^T > 1$.  Let $\delta_r = \delta(\boldsymbol{\theta}_r) \in \mathbb{R}$ and $\boldsymbol{\Sigma}_r^2$ be a $T \times T$ matrix that is independent of $n$. The approximate sampling distribution for $\hat{\boldsymbol{\delta}}^{_{(n)}} ~|~ \boldsymbol{\theta} = \boldsymbol{\theta}_{r}$ then takes the form
  \begin{equation}\label{eq:joint.mle}
   \mathcal{N}\left(\delta_r \times \mathbf{1}_T, \dfrac{1}{n}\times \boldsymbol{\Sigma}_r^2\right).
\end{equation}
\end{lemma} 

 \color{black}
 We prove Lemma \ref{lem0} and express $\boldsymbol{\Sigma}_r^2$ as a function of the inverse Fisher information in Appendix A.1 of the supplement. We only use the approximate sampling distribution in (\ref{eq:joint.mle}) for theoretical development and need not obtain $\boldsymbol{\Sigma}_r^2$ in practice. The entries in $\boldsymbol{\Sigma}_r^2$ implicitly account for the constants $c_t$ and $A_t$ specified for each analysis $t$. The result in Lemma \ref{lem0} is related to the joint canonical distribution  in sequential design theory \citep{jennison1997group, scharfstein1997semiparametric,jennison1999group}. \color{black}

Our proxies are also predicated on an asymptotic approximation to the posterior of $\delta = \delta(\boldsymbol{\theta})$ based on the Bernstein-von Mises (BvM) theorem outlined in Chapter 10.2 of \citet{vaart1998bvm}. The four conditions for the BvM theorem must therefore be satisfied to apply our methodology. The first three conditions concern the likelihood function and are also required for the asymptotic normality of the MLE.  The final condition concerns the analysis prior $p(\boldsymbol{\theta})$. This prior must be absolutely continuous with positive density in a neighborhood of the true parameter values $\boldsymbol{\theta}_r$ in iteration $r$. 
The limiting posterior of $\delta$ for the $t^{\text{th}}$ analysis prompted by the BvM theorem is
  \begin{equation}\label{eq:bvm}
  \mathcal{N}\left(\hat{\delta}^{_{(n)}}_{t,r}, \Sigma_{r, t, t}^2/n\right),
\end{equation}
where $\Sigma^2_{r, t, t}$ is the $(t,t)$-entry of $\boldsymbol{\Sigma}_r^2$ from (\ref{eq:joint.mle}). 


 To develop our proxies used for theoretical purposes, we use conditional cumulative distribution function (CDF) inversion to map realizations from the $T$-dimensional multivariate normal distribution in (\ref{eq:joint.mle}) to points $\boldsymbol{u} = \{u_t \}_{t=1}^T \in [0,1]^T$. For iteration $r$, we obtain the first component $\hat{\delta}^{_{(n)}}_{1, r}$ of the maximum likelihood estimate as the $u_{1}$-quantile of the sampling distribution of $\hat{\delta}^{_{(n)}}_{1} ~|~ \boldsymbol{\theta}_{r}$. For the remaining components, we obtain $\hat{\delta}^{_{(n)}}_{t, r}$ as the $u_{t}$-quantile of the sampling distribution of $\hat{\delta}^{_{(n)}}_{t} ~|~ \{\hat{\delta}^{_{(n)}}_{s} = \hat{\delta}^{_{(n)}}_{s, r}\}_{s=1}^{t-1}, \hspace*{0.1pt} \boldsymbol{\theta}_{r}$.

Implementing this process with $R$ points $\{\boldsymbol{u}_{r}\}_{r = 1}^R \sim \mathcal{U}\left([0,1]^{T}\right)$ and parameter values $\{\boldsymbol{\theta}_{r}\}_{r = 1}^R \sim \Psi$ gives rise to a sample from the approximate sampling distribution of $\hat{\boldsymbol{\delta}}^{_{(n)}}$ according to $\Psi$. For theoretical purposes, we substitute this sample $\{ \hat{\boldsymbol{\delta}}^{_{(n)}}_{ r}\}_{r=1}^R$ into the posterior approximation in (\ref{eq:bvm}) to yield a proxy sample of posterior probabilities. For each analysis $t$, we approximate the probability in the $t^{\text{th}}$ row of $\boldsymbol{\tau}(\mathcal{D}_{n})$ as
      \begin{equation}\label{eq:proxy}
\tau^{_{(n)}}_{t, r} = 
   \Phi\left(\dfrac{\delta_U - \hat{\delta}^{_{(n)}}_{t, r}}{\sqrt{\Sigma_{r, t, t}^2/n}}\right) - \Phi\left(\dfrac{\delta_L - \hat{\delta}^{_{(n)}}_{t, r}}{\sqrt{\Sigma_{r, t, t}^2/n}}\right)
\end{equation} 
where $\Phi(\cdot)$ is the standard normal CDF. The collection of $\{\tau^{_{(n)}}_{t, r}\}_{t = 1}^T$ values corresponding to $\{\boldsymbol{u}_{r}\}_{r = 1}^R \sim \mathcal{U}\left([0,1]^{T}\right)$ and $\{\boldsymbol{\theta}_{r}\}_{r = 1}^R \sim \Psi$ defines our proxy to the joint sampling distribution of $\boldsymbol{\tau}(\mathcal{D}_{n})$. Under the predictive approach, there are two sources of randomness in the proxy sampling distribution of $\boldsymbol{\tau}(\mathcal{D}_{n})$. The first source is associated with the parameter values $\boldsymbol{\theta}_{r}$ for iteration $r$. The second source is related to the point $\boldsymbol{u}_{r}$ used to generate the maximum likelihood estimate $\hat{\boldsymbol{\delta}}^{_{(n)}}_{ r}~|~\boldsymbol{\theta}_{r}$, which serves as a conduit for the data $\mathcal{D}_{n, r}$. 

When conditioning on particular values of $\boldsymbol{u}_{r}$ and $\boldsymbol{\theta}_{r}$, the value of $\tau^{_{(n)}}_{t, r}~|~\tau^{_{(n)}}_{t-1,r}$ is no longer a stochastic quantity. For $t = 1$, the conditioning set $\tau^{_{(n)}}_{t-1,r}$ is empty. We consider $\tau^{_{(n)}}_{t, r}$ conditional on $\tau^{_{(n)}}_{t-1,r}$ to model dependence in the joint proxy sampling distribution, and posterior summaries from previous stages do not provide additional information once $\tau^{_{(n)}}_{t-1,r}$ is known. Given values of $\boldsymbol{u}_{r}$ and $\boldsymbol{\theta}_{r}$, $\tau^{_{(n)}}_{t, r}~|~\tau^{_{(n)}}_{t-1,r}$ based on (\ref{eq:proxy}) is therefore a deterministic function of $n$. Lemma \ref{lem1} provides a standard structure for these deterministic functions under general conditions.

    \begin{lemma}\label{lem1}
     We suppose the conditions for Lemma \ref{lem0} are satisfied. Let the prior $p(\boldsymbol{\theta})$ satisfy the conditions for the BvM theorem. We consider a given point $\boldsymbol{u}_{r} \in [0,1]^{T}$, $\boldsymbol{\theta}_{r}$ value, and distribution for any covariates ${\bf{X}}$. For $t = 1, \dots, T$, the functions in (\ref{eq:proxy}) are such that 
    \begin{equation}\label{eq:lem1}
        \tau^{_{(n)}}_{t, r}~|~\tau^{_{(n)}}_{t-1,r} = \Phi\left(
           f_t(\delta_U, \boldsymbol{\theta}_{r})\sqrt{n} + g_t(\boldsymbol{u}_{r}) 
         \right) - \Phi\left(
           f_t(\delta_L, \boldsymbol{\theta}_{r})\sqrt{n} + g_t(\boldsymbol{u}_{r}) 
         \right),
    \end{equation} where $f_t(\cdot)$ and  $g_t(\cdot)$ are functions that do not depend on $n$.
\end{lemma} 

We prove Lemma \ref{lem1} and derive expressions for $\tau^{_{(n)}}_{t, r}~|~\tau^{_{(n)}}_{t-1,r}$ in Appendix A.2. 
In Section \ref{sec:methods.proxy2}, we use the result from (\ref{eq:lem1}) and extend the theory introduced here to create a proxy for the joint sampling distribution of posterior predictive probabilities in $\boldsymbol{\tau}_{\text{P}}(\mathcal{D}_{n})$. That proxy augments our proxy from this subsection to yield a proxy to the joint sampling distribution of posterior \emph{and} posterior predictive probabilities in $\boldsymbol{\tau}_{\hspace*{-1.25pt}*}(\mathcal{D}_{n})$.

\subsection{Proxies to Posterior Predictive Probabilities}\label{sec:methods.proxy2}

We now construct a proxy to the sampling distribution of $\boldsymbol{\tau}_{\text{P}}(\mathcal{D}_{n})$ that is also predicated on points $\{\boldsymbol{u}_{r}\}_{r = 1}^R \sim \mathcal{U}\left( [0,1]^T\right)$ and parameter values $\{\boldsymbol{\theta}_{r}\}_{r = 1}^R \sim \Psi$. 
To develop this proxy, we consider large-sample analogs for the components of the posterior predictive probability in (\ref{eq:ppp.def}). We illustrate how to create this proxy using $\tau_{\text{P},t}(\mathcal{D}_n)$ from (\ref{eq:ppp}), the posterior predictive probability at analysis $t < T$. We condition on the $n_t = c_tn$ already-observed observations at this analysis, and the final analysis would include $n_* = n_T - n_t = (c_T-c_t)n$ future observations. Our conduit for these $n_*$ observations generated from the posterior predictive distribution $ p_{\hspace*{0.75pt}\text{P}}(y~|~\mathcal{D}_{n_t})$ is the maximum likelihood estimate $\hat{\delta}^{_{(n)}}_{*,r}$. The conduit $\hat{\delta}^{_{(n)}}_{*,r}$ is asymptotically sufficient in that -- along with $\hat{\delta}^{_{(n)}}_{t,r}$, $n_t$, and $n_*$ -- it determines the limiting posterior distribution of $\delta$ at the final analysis based on the current and future data. We derive this limiting posterior distribution below.

In Appendix A.3 of the supplement, we show that an approximate sampling distribution for the corresponding MLE $\hat{\delta}^{_{(n)}}_{*,r}$ conditional on our conduit $\hat{\delta}^{_{(n)}}_{t,r}$ for the data $\mathcal{D}_{n_{t, r}}$ takes the form $\mathcal{N}(\hat{\delta}^{_{(n)}}_{t,r}, \kappa_{t,r}^2/n)$. \color{black} As detailed in Appendix A.3, the scaled variance $\kappa_{t,r}^2$ is a function of $\boldsymbol{\Sigma}^2_r$, $\{c_t\}_{t=1}^T$, and $\{A_t\}_{t=1}^T$ -- none of which depend on $n$ -- \color{black} under the simplifying assumption that the contribution to the limiting variance from the $n_*$ future observations is a function of the inverse Fisher information evaluated at $\boldsymbol{\theta} = \boldsymbol{\theta}_r$ instead of at $\boldsymbol{\theta} = \hat{\boldsymbol{\theta}}^{_{(n)}}_{t,r}$. This assumption is sensible because our theoretical proxies are based on large-sample results, and $\hat{\boldsymbol{\theta}}^{_{(n)}}_{t,r}$ should approximate $\boldsymbol{\theta}_r$ once the sample size is large enough to precisely identify the true parameter values. This assumption fails in settings with time-varying parameters, which often invalidate the regularity conditions for the asymptotic normality of the MLE.

The posterior distribution at the final analysis is based on a pooled sample of the initial data $\mathcal{D}_{n_{t, r}}$ and the $n_*$ future observations. Our large-sample analog for this pooling process creates a pooled MLE by combining $\hat{\delta}^{_{(n)}}_{t,r}$ and  $\hat{\delta}^{_{(n)}}_{*,r}$ using a weighted average. Based on the BvM theorem, the limiting posterior of $\delta$ at the final analysis is 
    \begin{equation}\label{eq:post.pooled}\delta~|~\hat{\delta}^{_{(n)}}_{t,r}, \hat{\delta}^{_{(n)}}_{*,r} \sim \mathcal{N}\left(\dfrac{c_t}{c_T}\hat{\delta}^{_{(n)}}_{t,r} + \dfrac{c_T-c_t}{c_T}\hat{\delta}^{_{(n)}}_{*,r}, \dfrac{\Sigma^2_{r, T, T}}{n}\right).
    \end{equation}
The posterior predictive probability in (\ref{eq:ppp.def}) conditions on the data available at analysis $t$ -- but not the future observations. The mean of the limiting posterior in (\ref{eq:post.pooled}) is thus a random quantity defined via the approximate distribution of $\hat{\delta}^{_{(n)}}_{*,r}~|~\hat{\delta}^{_{(n)}}_{t,r}$. Our large-sample analog to $\Pr\{\Pr(H_{1}~|~\mathcal{D}_{n_T}) \ge \gamma_{T}~|~\mathcal{D}_{n_{t, r}}\} $ involves quantiles of the limiting posterior of $\delta$ in (\ref{eq:post.pooled}). For any $q \in [0,1]$, the $q$-quantile of this posterior is also a random quantity:
    \begin{equation}\label{eq:quant}
\lambda_r(q) = \dfrac{c_t}{c_T}\hat{\delta}^{_{(n)}}_{t,r} + \dfrac{c_T-c_t}{c_T}\times \left( \hat{\delta}^{_{(n)}}_{t,r} + Z\dfrac{\kappa_{t,r}}{\sqrt{n}} \right) + \dfrac{\Sigma_{r, T, T}}{\sqrt{n}}\Phi^{-1}(q),
    \end{equation}
where $\hat{\delta}^{_{(n)}}_{*,r}$ has been expressed as a function of a standard normal random variable $Z$. 

    Our analog to the posterior predictive probability in (\ref{eq:ppp.def}) is the probability that $\lambda_r(q_L) > \delta_L$ and $\lambda_r(q_L + \gamma_T) < \delta_U$ for some $q_L \in [0, 1 - \gamma_T]$. For one-sided hypotheses, this value for $q_L$ does not depend on the sample size $n$: $q_L$ is respectively 0 and $1 - \gamma_T$ when $\delta_U$ is $\infty$ and $\delta_L$ is $-\infty$. In Appendix A.3 of the supplement, we show that the optimal value for $q_L \in [0, 1- \gamma_T]$ approaches a constant as $n \rightarrow \infty$ in the case where both $\delta_L$ and $\delta_U$ are finite. Since we only use our large-sample proxies for theoretical purposes, we regard $q_L$ as a constant that is independent of $n$. By rearranging (\ref{eq:quant}) to isolate for $Z$, we obtain the probability that $\lambda_r(q_L) > \delta_L$ and $\lambda_r(q_L + \gamma_T) < \delta_U$. This probability is our large-sample proxy to the probability in the $t^{\text{th}}$ row of $\boldsymbol{\tau}_{\text{P}}(\mathcal{D}_{n})$ for iteration $r$:
    \begin{equation}\label{eq:prob.ppp}
    \tau^{_{(n)}}_{\text{P}, t, r} = \Phi\left[\dfrac{\sqrt{n}(\delta_U - \hat{\delta}^{_{(n)}}_{t,r})}{\dfrac{\kappa_{t,r}(c_T-c_t)}{c_T}} - \dfrac{\Phi^{-1}(q_L + \gamma_T)\Sigma_{r, T, T}}{\dfrac{\kappa_{t,r}(c_T-c_t)}{c_T}} \right] - \Phi\left[\dfrac{\sqrt{n}(\delta_L - \hat{\delta}^{_{(n)}}_{t,r})}{\dfrac{\kappa_{t,r}(c_T-c_t)}{c_T}} - \dfrac{\Phi^{-1}(q_L)\Sigma_{r, T, T}}{\dfrac{\kappa_{t,r}(c_T-c_t)}{c_T}} \right].
    \end{equation}
    
We now reintroduce the variability associated with the available data $\mathcal{D}_{n_t, r}$ to construct a proxy to the joint sampling distribution of $\boldsymbol{\tau}_{\text{P}}(\mathcal{D}_{n})$. In Section \ref{sec:methods.proxy1}, we described how conduits $\{ \hat{\boldsymbol{\delta}}^{_{(n)}}_{ r}\}_{r=1}^R$ for the data could theoretically be mapped to points $\{\boldsymbol{u}_{r}\}_{r = 1}^R \sim \mathcal{U}\left([0,1]^T\right)$ and parameter values $\{\boldsymbol{\theta}_{r}\}_{r = 1}^R \sim \Psi$. The collection of $\{\tau^{_{(n)}}_{\text{P},t, r}\}_{t = 1}^T$ values corresponding to these points and parameter values comprises our proxy to the sampling distribution of $\boldsymbol{\tau}_{\text{P}}(\mathcal{D}_{n})$. Because our proxy to the sampling distribution of $\boldsymbol{\tau}(\mathcal{D}_{n})$ is defined using the \emph{same} points $\{\boldsymbol{u}_{r}\}_{r = 1}^R$ and parameter values $\{\boldsymbol{\theta}_{r}\}_{r = 1}^R$, we have constructed a proxy to the joint sampling distribution of posterior and posterior predictive probabilities in  $\boldsymbol{\tau}_{\hspace*{-1.25pt}*}(\mathcal{D}_{n})$. When conditioning on values of $\boldsymbol{u}_{r}$ and $\boldsymbol{\theta}_{r}$, $\tau^{_{(n)}}_{\text{P}, t, r}~|~\tau^{_{(n)}}_{t,r}$ based on (\ref{eq:prob.ppp}) is a deterministic function of $n$. Lemma \ref{lem2} provides a standard form for these functions. We prove this lemma and derive expressions for $\tau^{_{(n)}}_{\text{P}, t, r}~|~\tau^{_{(n)}}_{t,r}$ that account for dependence in the joint proxy sampling distribution in Appendix A.3.

        \begin{lemma}\label{lem2}
    We suppose the conditions for Lemma \ref{lem1} are satisfied. We consider a given point $\boldsymbol{u}_{r} \in [0,1]^{T}$, $\boldsymbol{\theta}_{r}$ value, and distribution for any covariates ${\bf{X}}$. For $t = 1, \dots, T-1$, the functions in (\ref{eq:prob.ppp}) are such that 
     \begin{equation}\label{eq:lem2}
     \tau^{_{(n)}}_{\text{P}, t, r}~|~\tau^{_{(n)}}_{t,r} = \Phi\left(
           f_{\text{P}, t}(\delta_U, \boldsymbol{\theta}_{r})\sqrt{n} + g_{\text{P}, t}(\boldsymbol{u}_{r}, q_L + \gamma_T) 
         \right) - \Phi\left(
           f_{\text{P}, t}(\delta_L, \boldsymbol{\theta}_{r})\sqrt{n} + g_{\text{P}, t}(\boldsymbol{u}_{r}, q_L) 
         \right),
    \end{equation}
    where $f_{\text{P}, t}(\cdot)$ and  $g_{\text{P}, t}(\cdot)$ are functions that do not depend on $n$.
\end{lemma} 

    Our proxy to the sampling distribution of $\boldsymbol{\tau}_{\hspace*{-1.25pt}*}(\mathcal{D}_{n})$ relies on asymptotic results, so it may differ materially from the true sampling distribution for finite $n$. Therefore, this proxy only motivates our theoretical result in Theorem \ref{thm1}, which utilizes the deterministic functions derived in Lemmas \ref{lem1} and \ref{lem2}. Theorem \ref{thm1} guarantees that the logits of $\tau^{_{(n)}}_{t, r}~|~\tau^{_{(n)}}_{t-1,r}$ and $\tau^{_{(n)}}_{\text{P}, t, r}~|~\tau^{_{(n)}}_{t,r}$ are approximately linear functions of $n$ for all $t \in \{1, \dots, T\}$. We later adapt this result to estimate the error probabilities of GSDs across a wide range of sample sizes by estimating the true sampling distribution of $\boldsymbol{\tau}_{\hspace*{-1.25pt}*}(\mathcal{D}_{n})$ at only two values of $n$.

    \begin{theorem}\label{thm1}
     We suppose the conditions for Lemma \ref{lem1} are satisfied. Define $\emph{logit}(x) = \emph{log}(x) - \emph{log}(1-x)$. We consider a given point $\boldsymbol{u}_{r} \in [0,1]^{T}$, $\boldsymbol{\theta}_{r}$ value, and distribution for any covariates ${\bf{X}}$. The functions $\{\tau^{_{(n)}}_{t, r}~|~\tau^{_{(n)}}_{t-1,r}\}_{t=1}^T$ based on (\ref{eq:lem1}) and $\{\tau^{_{(n)}}_{\text{P}, t, r}~|~\tau^{_{(n)}}_{t,r}\}_{t=1}^{T-1}$ based on (\ref{eq:lem2}) are such that
 \begin{enumerate}
     \item[(a)] $\lim\limits_{n \rightarrow \infty} \dfrac{d}{dn}~\emph{logit}\left(\tau^{_{(n)}}_{t, r}~|~\tau^{_{(n)}}_{t-1,r}\right)= (0.5 - \mathbb{I}\{\delta_{r} \notin (\delta_{L}, \delta_{U})\})\times\emph{min}\{f_t(\delta_{U}, \boldsymbol{\theta}_{r})^2, f_t(\delta_{L}, \boldsymbol{\theta}_{r})^2\} $. 
     \item[(b)] $\lim\limits_{n \rightarrow \infty} \dfrac{d}{dn}~\emph{logit}\left(\tau^{_{(n)}}_{\text{P}, t, r}~|~\tau^{_{(n)}}_{t,r}\right)= (0.5 - \mathbb{I}\{\delta_{r} \notin (\delta_{L}, \delta_{U})\})\times\emph{min}\{f_{\text{P},t}(\delta_{U}, \boldsymbol{\theta}_{r})^2, f_{\text{P},t}(\delta_{L}, \boldsymbol{\theta}_{r})^2\} $. 
 \end{enumerate}
\end{theorem} 

Theorem \ref{thm1} is novel to this paper; we prove parts $(a)$ and $(b)$ in Appendix B of the supplement. We now discuss the practical implications of this theorem. The limiting derivatives in parts $(a)$ and $(b)$ are constants that do not depend on $n$. Moreover, these limiting derivatives do not depend on the point $\boldsymbol{u}_{r}$ which controls the dependence within the joint proxy sampling distribution of $\boldsymbol{\tau}_{\hspace*{-1.25pt}*}(\mathcal{D}_{n})$. The limiting derivatives of (i) $\text{logit}(\tau^{_{(n)}}_{t, r})$ and $\text{logit}(\tau^{_{(n)}}_{t, r}~|~\tau^{_{(n)}}_{t-1,r})$ and (ii) $\text{logit}(\tau^{_{(n)}}_{\text{P}, t, r})$ and $\text{logit}(\tau^{_{(n)}}_{\text{P}, t, r}~|~\tau^{_{(n)}}_{t,r})$ are therefore the same. In the joint \emph{proxy} sampling distribution, the linear approximations to $l^{_{(n)}}_{t, r} = \text{logit}(\tau^{_{(n)}}_{t, r}~|~\tau^{_{(n)}}_{t-1,r})$ and $l^{_{(n)}}_{\text{P},t, r} = \text{logit}(\tau^{_{(n)}}_{\text{P},t, r}~|~\tau^{_{(n)}}_{t,r})$ as functions of $n$ are thus good global approximations for large sample sizes. These linear approximations should be locally suitable for a range of smaller sample sizes. 

Under the conditional approach where $\{\boldsymbol{\theta}_{r}\}_{r=1}^R$ are the same, the (conditional) quantiles of the sampling distributions of $l^{_{(n)}}_{t, r}$ and $l^{_{(n)}}_{\text{P},t, r}$ therefore change linearly as a function of $n$. In Sections \ref{sec:stop} and \ref{sec:power}, we adapt these linear trends from the proxy sampling distribution in the absence of stopping rules to flexibly model the logits of $\boldsymbol{\tau}_{\hspace*{-1.25pt}*}(\mathcal{D}_{n})$ as linear functions of $n$ when independently simulating samples $\mathcal{D}_{n, r}$ according to $\boldsymbol{\theta}_{r} \sim \Psi$ under the conditional or predictive approach for a design with early stopping.  Although the proxy sampling distribution is predicated on asymptotic results for the first analysis, we illustrate the good performance of our design procedure with finite sample sizes $n$ in Section \ref{sec:ex}.

   \section{Stopping Rules}\label{sec:stop}

   We now introduce stopping rules based on posterior summaries that inform interim decisions in GSDs \emph{with} early stopping. We first consider stopping rules based on the posterior probabilities in (\ref{eq:pp}). Early stopping may be facilitated  by comparing $\boldsymbol{\tau}(\mathcal{D}_{n})$ to success thresholds $\boldsymbol{\gamma}=\{\gamma_t\}_{t=1}^T \in [0,1]^T$ or failure thresholds $\boldsymbol{\xi}=\{\xi_t\}_{t=1}^T < \{\gamma_t\}_{t=1}^T$. If not already stopped in a previous stage, the experiment may be stopped for success at analysis $t$ if $\tau_t(\mathcal{D}_{n}) \ge \gamma_t$, or it could be stopped for failure if $\tau_t(\mathcal{D}_{n}) < \xi_t$. We next consider stopping rules based on the posterior predictive probabilities in (\ref{eq:ppp}). Such early stopping can be implemented using success thresholds $\boldsymbol{\eta} = \{\eta_t\}_{t=1}^{T-1} \in [0,1]^{T-1}$ and failure thresholds $\boldsymbol{\rho} = \{\rho_t\}_{t=1}^{T-1} < \{\eta_t\}_{t=1}^{T-1}$. If not stopped in a previous stage, the experiment may be respectively stopped for success or failure at analysis $t$ if $\tau_{\text{P}, t}(\mathcal{D}_{n}) \ge \eta_t$ or $\tau_{\text{P}, t}(\mathcal{D}_{n}) < \rho_t$.

   To consider the error probabilities of sequential designs with general stopping rules, we define variables \color{black}  $\{\nu_t(\mathcal{D}_{n})\}_{t=1}^T$ that denote the status of the experiment after the $t^{\text{th}}$ analysis:
   $$
   \nu_t(\mathcal{D}_n) = \begin{cases} -1 & \text{if stopped for failure before the end of stage \emph{t}} \\
   0 & \text{if continuing beyond stage \emph{t}} \\
   1 & \text{if stopped for success before the end of stage \emph{t}}.
       
   \end{cases} $$
   We suppose $\nu_T(\mathcal{D}_n) \in \{-1, 1\}$ such that the experiment stops for failure at the final analysis if it does not stop for success. \color{black} An example design with $T=2$ analyses underscores the relationship between $\boldsymbol{\tau}_{\hspace*{-1.25pt}*}(\mathcal{D}_{n})$ and $\nu_t(\mathcal{D}_{n})$. This example design considers stopping for success based on $\boldsymbol{\tau}(\mathcal{D}_{n})$ before considering stopping for failure based on $\boldsymbol{\tau}_{\text{P}}(\mathcal{D}_{n})$. We have that $\nu_1(\mathcal{D}_{n}) = 1$ if $\tau_1(\mathcal{D}_{n}) \ge \gamma_1$,  $\nu_1(\mathcal{D}_{n}) = -1$ if $\tau_1(\mathcal{D}_{n}) < \gamma_1$ and $\tau_{\text{P},1}(\mathcal{D}_{n}) < \rho_1$, and $\nu_1(\mathcal{D}_{n}) = 0$ if $\tau_1(\mathcal{D}_{n}) < \gamma_1$ and $\tau_{\text{P},1}(\mathcal{D}_{n}) \ge \rho_1$. Moreover, $\nu_2(\mathcal{D}_{n}) = 1$ if $\nu_1(\mathcal{D}_{n}) = 1$ or $\nu_1(\mathcal{D}_{n}) = 0$ and $\tau_2(\mathcal{D}_{n}) \ge \gamma_2$; $\nu_2(\mathcal{D}_{n}) = -1$ if $\nu_1(\mathcal{D}_{n}) = -1$ or $\nu_1(\mathcal{D}_{n}) = 0$ and $\tau_2(\mathcal{D}_{n}) < \gamma_2$.

We now define error probabilities with respect to the model from which $\boldsymbol{\theta}$ values are drawn. For a given model $\Psi$, the probability of stopping for success before the end of stage $t$ is
  \begin{equation}\label{eq:doc}
  \mathbb{E}_{\Psi}[\Pr(\nu_t(\mathcal{D}_{n}) = \dagger~|~\boldsymbol{\theta})] = \int \Pr(\nu_t(\mathcal{D}_{n}) = \dagger~|~\boldsymbol{\theta})\Psi(\boldsymbol{\theta})d\boldsymbol{\theta},
\end{equation} 
where $\dagger = 1$. The corresponding probability of stopping for failure before the end of stage $t$ is given by (\ref{eq:doc}) when $\dagger = -1$. Given the simulation results from Section \ref{sec:prelim}, these probabilities in (\ref{eq:doc}) for analysis $t$ can be estimated using 
  \begin{equation}\label{eq:power}
  \dfrac{1}{R}\sum_{r=1}^R \mathbb{I}\left\{\nu_t(\mathcal{D}_{n, r}) = \dagger\right\},
\end{equation} 
where $\dagger \in \{-1, 1\}$ and $\mathcal{D}_{n, r}$ are generated using $\boldsymbol{\theta}_r$ obtained via $\Psi$.  The power to correctly stop for success before the end of the experiment (and not make a type II error) is $\mathbb{E}_{\Psi_1}[\Pr(\nu_T(\mathcal{D}_{n}) = 1~|~\boldsymbol{\theta})]$, where $\Psi_1$ is a probability model such that $H_1$ is true. We estimate power using (\ref{eq:power}) with $t = T$ and $\dagger = 1$ when $\{\mathcal{D}_{n_t, r}\}_{t = 1}^T$ are generated using $\boldsymbol{\theta}_r$ obtained via $\Psi_1$. 

The type I error rate related to incorrectly stopping for success at any analysis is $\mathbb{E}_{\Psi_0}[\Pr(\nu_T(\mathcal{D}_{n}) = 1~|~\boldsymbol{\theta})]$ where $\Psi_0$ is a probability model such that $H_0$ is true. Using $\Psi_0$ instead of $\Psi_1$, this probability can be estimated as in (\ref{eq:power}) with $t = T$ and $\dagger = 1$. \color{black} We could also use (\ref{eq:power}) with $\dagger = 1$ to estimate the cumulative probability of incorrectly stopping for success under $\Psi_0$ at an analysis $t < T$ if strictly calibrating the type I error rate to a desired alpha-spending function. Furthermore, we could use (\ref{eq:power}) with $\dagger = -1$ to estimate the cumulative probability of incorrectly stopping for failure under $\Psi_1$ at an analysis $t < T$ if calibrating the type II error rate to a specified beta-spending function. \color{black} 

We emphasize that operating characteristics can be computed to appropriately account for stopping criteria using an estimate of the joint sampling distribution $\{\boldsymbol{\tau}_{\hspace*{-1.25pt}*}(\mathcal{D}_{n, r})\}_{r=1}^R$ that applies in the absence of stopping rules.  In iteration $r$, suppose the $t^{\text{th}}$ analysis is the first one at which $\nu_t(\mathcal{D}_{n, r}) \in \{-1,1\}$. We can then simply ignore the rows in $\boldsymbol{\tau}_{\hspace*{-1.25pt}*}(\mathcal{D}_{n, r})$ that correspond to analyses $t + 1, \dots, T$. This use of $\{\boldsymbol{\tau}_{\hspace*{-1.25pt}*}(\mathcal{D}_{n, r})\}_{r=1}^R$ from a design without early stopping to compute error probabilities is suitable under the following conditions. First, the proportion of observations in each stage of the design must be pre-specified (as enforced by the constraint $\{n_t\}_{t=1}^T = n \times \{c_t\}_{t=1}^T$ for constants $c_1=1$ and $\{c_t\}_{t=2}^T > 1$). Second, the allocation probabilities $\{A_t\}_{t=1}^T$ for each stage of the design must also be pre-specified.

\color{black} 
The decision thresholds $\boldsymbol{\gamma}$, $\boldsymbol{\eta}$, $\boldsymbol{\xi}$ and $\boldsymbol{\rho}$ bound the type I error rates and interim type II error rates of sequential designs. \color{black} The sample sizes $\{n_t\}_{t=1}^T$ are selected to ensure the experiment has a small enough type II error rate at the final analysis (i.e., to guarantee power is sufficiently large). In Section \ref{sec:power}, we present a streamlined method to select sample sizes and decision thresholds for GSDs with early stopping.

   \section{Efficient Design of Group Sequential Experiments}\label{sec:power}

   We  generalize  the  results  from  Theorem  \ref{thm1}  to  develop  a  procedure  to  choose  sample  sizes and tune decision thresholds for Bayesian GSDs  with early stopping in Algorithm \ref{alg1}. This procedure requires that we estimate the sampling distribution of posterior summaries $\boldsymbol{\tau}_{\hspace*{-1.25pt}*}(\mathcal{D}_n)$  in the absence of stopping rules  by simulating data $\mathcal{D}_n$ at \emph{only} two values of $n$: $n_{a}$ and $n_b$. The initial sample size for the first analysis $n_a$ can be selected based on the anticipated budget for the sequential experiment. In Algorithm \ref{alg1}, we add a subscript to $\mathcal{D}_{n,r}$ between $n$ and $r$ that distinguishes whether the data are generated according to the model $\Psi_0$ or $\Psi_1$ defined in Section \ref{sec:stop}. In addition to the choices discussed previously, we specify a distribution with parameters $\boldsymbol{\zeta}$ for any potential covariates ${\bf{X}}_{n_t \times w}$. 

   \color{black}
   We also define criteria for the error probabilities. In Algorithm \ref{alg1}, we calibrate the cumulative type II error rates, $\{\mathbb{E}_{\Psi_1}[\Pr(\nu_t(\mathcal{D}_{n}) = -1~|~\boldsymbol{\theta})]\}_{t=1}^T$, to a beta-spending function $\boldsymbol{\beta} = (\beta_1, \dots, \beta_T)$. %
   We also calibrate the cumulative type I error rates, $\{\mathbb{E}_{\Psi_0}[\Pr(\nu_t(\mathcal{D}_{n}) = 1~|~\boldsymbol{\theta})]\}_{t=1}^T$, to an alpha-spending function $\boldsymbol{\alpha} = (\alpha_1, \dots, \alpha_T)$. 
   While our methodology can be used with general alpha- and beta-spending functions, these functions may be chosen to minimize the expected sample size of the GSD or attain other optimality criteria (see e.g., \citet{jennison1987efficient, jennison1999group}). \color{black} Algorithm \ref{alg1} details an application of our methodology with the conditional approach, and we later describe potential modifications.

         \begin{algorithm}
\caption{Procedure to Determine Sample Sizes and Decision Thresholds}
\label{alg1}

\begin{algorithmic}[1]
\setstretch{1.2}
\Procedure{Design}{$p(y~|~ \boldsymbol{\theta})$, $\delta(\cdot)$, $\delta_L$, $\delta_U$, $p(\boldsymbol{\theta})$, $\Psi_0$, $\Psi_1$, $R$, $M$, $n_a$, $\{c_t\}_{t=1}^T$, $\{A_t\}_{t=1}^T$, $\boldsymbol{\zeta}$, $\boldsymbol{\alpha}$, $\boldsymbol{\beta}$}
\State  Compute $\{\boldsymbol{\tau}_{\hspace*{-1.25pt}*}(\mathcal{D}_{n_a, 0, r})\}_{r = 1}^R$ obtained with $\boldsymbol{\theta}_r \sim \Psi_0$. 
\State  Compute $\{\boldsymbol{\tau}_{\hspace*{-1.25pt}*}(\mathcal{D}_{n_a, 1, r})\}_{r = 1}^R$ obtained with $\boldsymbol{\theta}_r \sim \Psi_1$. 
\State \color{black} Specify thresholds from $\boldsymbol{\gamma}$, $\boldsymbol{\xi}$, $\boldsymbol{\eta}$, and $\boldsymbol{\rho}$ to calibrate $\{R^{-1}\sum_{r=1}^R\mathbb{I}\{\nu_t(\mathcal{D}_{n_a, 0, r}) = 1\}\}_{t=1}^T$  to $\{\alpha_t\}_{t=1}^T$ and \linebreak \hspace*{12pt} $\{R^{-1}\sum_{r=1}^R\mathbb{I}\{\nu_t(\mathcal{D}_{n_a, 1, r}) = -1\}\}_{t=1}^{T-1}$ to $\{\beta_t\}_{t=1}^{T-1}$. \color{black}
\State If $R^{-1}\sum_{r=1}^R\mathbb{I}\{\nu_T(\mathcal{D}_{n_a, 1, r}) = 1\} \ge 1 - \beta_T$, choose $n_b < n_a$. If not, choose $n_b > n_a$.
\State  \color{black} Compute $\{\boldsymbol{\tau}_{\hspace*{-1.25pt}*}(\mathcal{D}_{n_b, 0, r})\}_{r = 1}^R$ obtained with $\boldsymbol{\theta}_r \sim \Psi_0$. \color{black}
\State  Compute $\{\boldsymbol{\tau}_{\hspace*{-1.25pt}*}(\mathcal{D}_{n_b, 1, r})\}_{r = 1}^R$ obtained with $\boldsymbol{\theta}_r \sim \Psi_1$. \color{black}
\For{$j$ in \{0,1\}} \color{black}
\For{$d$ in $1$:$R$}
 \For{$t$ in $1$:$T$}
 \State Let $\mathcal{D}_{n_a, j, r}$ correspond to the $d^{\text{th}}$ order statistic of $\{l_t(\mathcal{D}_{n_a, j, r})\}_{r = 1}^R$. 
\State Pair the $d^{\text{th}}$ order statistics of $\{l_t(\mathcal{D}_{n_a, j, r})\}_{r = 1}^R$ and $\{l_t(\mathcal{D}_{n_b, j, r})\}_{r = 1}^R$ with   linear  approxima- \linebreak \hspace*{58pt} tions  to obtain $\hat{l}_t(\mathcal{D}_{n, j, r})$ estimates for new $n$ values.
\If {$t < T$}
 \State Repeat Lines 9 and 10 with $\{l_{\text{P}, t}(\mathcal{D}_{n_a, j, r})\}_{r = 1}^R$ and $\{l_{\text{P}, t}(\mathcal{D}_{n_b, j, r})\}_{r = 1}^R$ to  estimate \linebreak \hspace*{75pt} $\hat{l}_{\text{P}, t}(\mathcal{D}_{n, j, r})$ for new $n$.
 \EndIf
 \EndFor
 \EndFor
 \State Obtain $\{\hat{\boldsymbol{\tau}}_{\hspace*{-1.25pt}*}(\mathcal{D}_{n, j, r})\}_{r = 1}^R$ as the expits of  $\{\hat{l}_t(\mathcal{D}_{n, j, r})\}_{t = 1}^{T}$ and  $\{\hat{l}_{\text{P}, t}(\mathcal{D}_{n, j, r})\}_{t = 1}^{T-1}$.
 \EndFor
\State Find $n_c$, the smallest $n \in \mathbb{Z}^+$ such that $R^{-1}\sum_{r=1}^R\mathbb{I}\{\hat{\nu}_T(\mathcal{D}_{n, 1, r}) = 1\} \ge 1 - \beta_T$  \color{black} while  recalibrating the \linebreak \hspace*{12pt} thresholds from $\boldsymbol{\gamma}$, $\boldsymbol{\xi}$, $\boldsymbol{\eta}$, and $\boldsymbol{\rho}$ as in Line 4 for each $n$. \color{black}
\State \Return $n_c$ as recommended $n$ and $\{\boldsymbol{\gamma}, \boldsymbol{\xi}, \boldsymbol{\eta}, \boldsymbol{\rho}\}$ as recommended decision thresholds

\EndProcedure

\end{algorithmic}
\end{algorithm}

We now elaborate on several steps of Algorithm \ref{alg1}. \color{black} In Line 4, we choose suitable vectors for the relevant decision thresholds to approximately calibrate the type I and interim type II error rates to $\boldsymbol{\alpha}$ and $\boldsymbol{\beta}$. We propose the following process as one option to tune the decision thresholds. We first use binary search to select $\gamma_1$ or $\eta_1$ as the smallest success threshold such that $R^{-1}\sum_{r=1}^R\mathbb{I}\{\nu_1(\mathcal{D}_{n_a, 0, r}) = 1\} \le \alpha_1$. For analysis $t$, $\gamma_t$ or $\eta_t$ are then similarly selected to ensure $R^{-1}\sum_{r=1}^R\mathbb{I}\{\nu_t(\mathcal{D}_{n_a, 0, r}) = 1\} \le \alpha_t$ given the previously tuned success thresholds for the earlier analyses. Given these tuned success thresholds, the failure thresholds $\xi_t$ or $\rho_t$ are iteratively chosen for analyses $t = 1$ to $T-1$ as the largest failure thresholds such that $R^{-1}\sum_{r=1}^R\mathbb{I}\{\nu_t(\mathcal{D}_{n_a, 1, r}) = -1\} \le \beta_t$. 

These tuned decision thresholds approximately maintain the desired cumulative type II error rates. Because the success thresholds are tuned before accounting for stopping for failure, the resulting cumulative type I error rates may be slightly lower than desired (see Section \ref{sec:ex} for an illustration). We use the tuning procedure proposed here since it cannot lead to a type I error rate that is higher than intended. While it is possible to use our methods while jointly tuning the success and failure thresholds, such procedures -- which are typically more computationally intensive or adhoc -- are not considered in this work. \color{black}

All posterior summaries approximated in Lines 2 to 7 of Algorithm \ref{alg1} are obtained by simulating data given parameter values from $\Psi_0$ or $\Psi_1$. We compute logits of these summaries under $\Psi_j$ for $j \in \{0, 1\}$: $l_t(\mathcal{D}_{n, j, r}) = \text{logit}(\tau_t(\mathcal{D}_{n, j, r}))$ and $l_{\text{P}, t}(\mathcal{D}_{n, j, r}) = \text{logit}(\tau_{\text{P}, t}(\mathcal{D}_{n, j, r}))$. If not all components of $\boldsymbol{\tau}_{\hspace*{-1.25pt}*}(\mathcal{D}_{n})$ inform decision rules in a particular design, various rows in this vector may be ignored. We recommend calculating posterior summaries using nonparametric kernel density estimates so that these logits are finite. 

We construct linear approximations separately for each summary using logits corresponding to sample sizes $n_a$ and $n_b$ under the model $\Psi_j$ in Lines 8 to 15. This second sample size $n_b$ could be chosen in light of whether power at $n_a$ in Line 5 is sufficiently large. Alternatively, $n_b$ could be chosen to explore a range of relevant sample sizes between $n_a$ and $n_b$. We elaborate on the choice of $n_a$ and $n_b$ for our examples in Section \ref{sec:ex}. We use these linear approximations to estimate logits of posterior summaries under $\Psi_j$ for new values of $n$ as $\hat{l}_t(\mathcal{D}_{n, j, r})$ or $\hat{l}_{\text{P}, t}(\mathcal{D}_{n, j, r})$. We place a hat over the $l$ here to convey that this logit was estimated using a linear approximation instead of a sample of data. To maintain the proper level of dependence in the joint sampling distribution of $\boldsymbol{\tau}_{\hspace*{-1.25pt}*}(\mathcal{D}_{n})$, we group the linear functions from Lines 12 and 14 across all posterior summaries (separately for each $\Psi_j$) based on the sample $\mathcal{D}_{n_a, j, r}$ that defined the linear approximations.

Given the linear trend in the proxy sampling distribution quantiles discussed in Section \ref{sec:methods.proxy2}, these linear approximations can be constructed based on order statistics of estimates of the true sampling distributions under the conditional approach. Because Theorem \ref{thm1} ensures that the limiting slopes of (i) $\text{logit}(\tau^{_{(n)}}_{t, r})$ and $\text{logit}(\tau^{_{(n)}}_{t, r}~|~\tau^{_{(n)}}_{t-1,r})$ and (ii) $\text{logit}(\tau^{_{(n)}}_{\text{P}, t, r})$ and $\text{logit}(\tau^{_{(n)}}_{\text{P}, t, r}~|~\tau^{_{(n)}}_{t,r})$ from the proxy sampling distribution are the same, we use the \emph{marginal} sampling distributions for each row of $\boldsymbol{\tau}_{\hspace*{-1.25pt}*}(\mathcal{D}_n)$ to estimate slopes for the \emph{conditional} logits. Under the predictive approach for $\Psi_0$ or $\Psi_1$, the process in Lines 8 to 15 can be modified. We split the logits of the posterior summaries for each $n$ value into subgroups based on the order statistics of their $\delta_{r}$ values before constructing the linear approximations. Throughout this paper, we split the logits into 10 bins under the predictive approach.

In Line 16, we find the smallest value of $n$ such that power estimated using the indicators $\{\hat{\boldsymbol{\nu}}_T(\mathcal{D}_{n, 1, r})\}_{r = 1}^R$ corresponding to $\{\hat{\boldsymbol{\tau}}_{\hspace*{-1.25pt}*}(\mathcal{D}_{n, 1, r})\}_{r = 1}^R$ is at least $1 - \beta_T$. \color{black} For each $n$ considered, the decision thresholds are recalibrated  using the process described above with the sampling distribution estimates $\{\hat{\boldsymbol{\tau}}_{\hspace*{-1.25pt}*}(\mathcal{D}_{n, 0, r})\}_{r = 1}^R$ and $\{\hat{\boldsymbol{\tau}}_{\hspace*{-1.25pt}*}(\mathcal{D}_{n, 1, r})\}_{r = 1}^R$ constructed using our linear approximations. Sample sizes throughout the design are obtained using the constants $\{c_t\}_{t=2}^T$.  
A simpler version of Algorithm \ref{alg1} that only requires calibration of the cumulative type I and II error rates at analysis $T$ is provided in Appendix C of the supplement. \color{black}

\color{black}
    \section{Performance for Example Designs}\label{sec:ex}

    \subsection{Clinical Trial with Dynamic Borrowing}\label{sec:ex.1}

    \color{black}

  We assess the performance of Algorithm \ref{alg1} with example designs based on the Canadian placebo-controlled randomized trial of tecovirimat in non-hospitalized patients with clade II Mpox (PLATINUM-CAN) \citep{klien2024tecovirimat}. This trial employs a fixed design with a single frequentist analysis at the end of the trial. For illustration, we consider various Bayesian GSDs with $T = 3$ potential analyses. The main goal of the trial is to establish whether the antiviral drug tecovirimat reduces the duration of illness associated with Mpox infection. 
  One additional outcome for this trial denotes whether all active lesions for a patient had resolved after 14 days of treatment. \color{black} Historical data are available on this outcome from a clade I Mpox observational study \citep{pittman2023clinical} that can inform the placebo arm parameters. Therefore, we use this binary outcome for trial design in this subsection to present a scenario with dynamic borrowing. \color{black} As this outcome is not the primary outcome in PLATINUM-CAN, our conclusions are not directly applicable to the real trial.
  
  We suppose the binary outcomes in the tecovirimat and placebo arms respectively come from $\text{BIN}(1, \theta_{\text{TEC}})$ and $\text{BIN}(1, \theta_{\text{PL}})$ distributions. The target of inference $\delta(\boldsymbol{\theta})$ is the difference between the lesion resolution rates in the two arms: $\theta_{\text{TEC}} - \theta_{\text{PL}}$. The hypotheses for this trial are $H_0 : \delta \le 0$ vs.\ $H_{1}: \delta > 0$. That is, $\delta_L = 0$ and $\delta_U = \infty$. Our five example designs -- D1, D2, D3, D4, and D5 -- consider different choices for the prior distribution $p(\boldsymbol{\theta})$ and probability models $\Psi_0$ and $\Psi_1$. All designs have equally spaced interim analyses such that $(c_1, c_2, c_3) = (1, 2, 3)$. For all designs, the allocation ratio for tecovirimat to placebo is 2:1 in stage 1 and 1:1 in stages 2 and 3. D1 is our primary design, and we consider the others for sensitivity analysis. 

  All designs accommodate stopping for success based on $\boldsymbol{\tau}(\mathcal{D}_n)$. We tune the success thresholds $\boldsymbol{\gamma}$ to approximately attain a cumulative alpha-spending vector, $\boldsymbol{\alpha} = (0.0013, 0.0061, 0.0250)$, which corresponds to a Hwang-Shih-DeCani spending function \citep{hwang1990group} with family-wise error rate (FWER) of 0.025. D1 to D4 also accommodate early stopping for failure based on $\boldsymbol{\tau}_{\text{P}}(\mathcal{D}_n)$. Given the previously tuned $\boldsymbol{\gamma}$, we tune the failure thresholds $\boldsymbol{\rho}$ to approximately attain a cumulative beta-spending vector, $\boldsymbol{\beta} = (0.01, 0.05, 0.15)$. The beta-spending sequence for D3 is instead $\boldsymbol{\beta} = (0, 0, 0.15)$. For all designs, the sample size $n$ is thus selected to attain a target power of 0.85. 

  A diffuse $\text{BETA}(1,1)$ prior is assigned to $\theta_{\text{TEC}}$ in all designs. For all designs except D3, we use a robust meta-analytic prior proposed by \citet{schmidli2014robust} to incorporate historical information for the placebo arm from the clade I Mpox observational study in \citet{pittman2023clinical}. The resulting prior distribution for $\theta_{\text{PL}}$ is a mixture distribution with two components: a $\text{BETA}(134, 53)$ component informed by the historical information and a diffuse $\text{BETA}(1, 1)$ component. At an arbitrary analysis, suppose $n_{\text{PL}}$ cumulative patients have been randomized to the placebo arm, with $y_{\text{PL}}$ patients experiencing lesion resolution after 14 days of treatment. The subsequent posterior distribution for $\theta_{\text{PL}}~|~y_{\text{PL}}, n_{\text{PL}}$ is a mixture of the $\text{BETA}(134 + y_{\text{PL}}, 53 + n_{\text{PL}}- y_{\text{PL}})$ and $\text{BETA}(1 + y_{\text{PL}}, 1 + n_{\text{PL}}- y_{\text{PL}})$ distributions, where the posterior weights for the two components are updated using marginal likelihoods based on the observed binary data. In particular, the informative component of the prior is dynamically downweighted if it does not align with the observed trial data from the clade II Mpox patients. When the parameters $\boldsymbol{\theta}_r$ for data generation are fixed, the posterior weight for each component of this prior approaches a constant as $n \rightarrow \infty$ \citep{schmidli2014robust}. Our theoretical results in Section \ref{sec:methods} and design procedure in Algorithm \ref{alg1} can thus be applied. In D3, a $\text{BETA}(1,1)$ prior is also assigned to $\theta_{\text{PL}}$ for illustration.

  For D1, D3, and D5, $\Psi_0$ is such that the outcomes for both arms are generated via $\text{BIN}(1, 0.7)$ distributions; $\Psi_1$ is such that the outcomes for the placebo and tecovirimat arms respectively come from $\text{BIN}(1, 0.7)$ and $\text{BIN}(1, 0.8)$ distributions. That is, $\Psi_0$ and $\Psi_1$ are defined under the conditional approach. In D4, $\Psi_0$ is the same as in D1, D3, and D5, but $\Psi_1$ is defined under the predictive approach such that $\{\delta(\boldsymbol{\theta}_r)\}_{r=1}^R$ come from a $\mathcal{N}(0.1, 0.03^2)$ distribution that is truncated between 0.05 and 0.15. The probability models for D1, D3, D4, and D5 are chosen to reflect scenarios where the historical information improves power without greatly inflating the type I error rate.

  In D2, $\Psi_0$ and $\Psi_1$ are defined under the conditional approach such that the outcomes for the placebo and tecovirimat arms respectively come from $\text{BIN}(1, 0.75)$ and $\text{BIN}(1, 0.843)$ distributions. The average standardized effect sizes are thus similar across all designs. D2 is defined such that the historical information introduces negative bias into the posterior distribution of $\theta_{\text{PL}}$. As a result, positive bias is introduced into the posterior distribution of $\delta$, and we require more conservative decision thresholds to control the type I error rate.

\begin{figure}[!tb] \centering 
		\includegraphics[width = 0.8\textwidth]{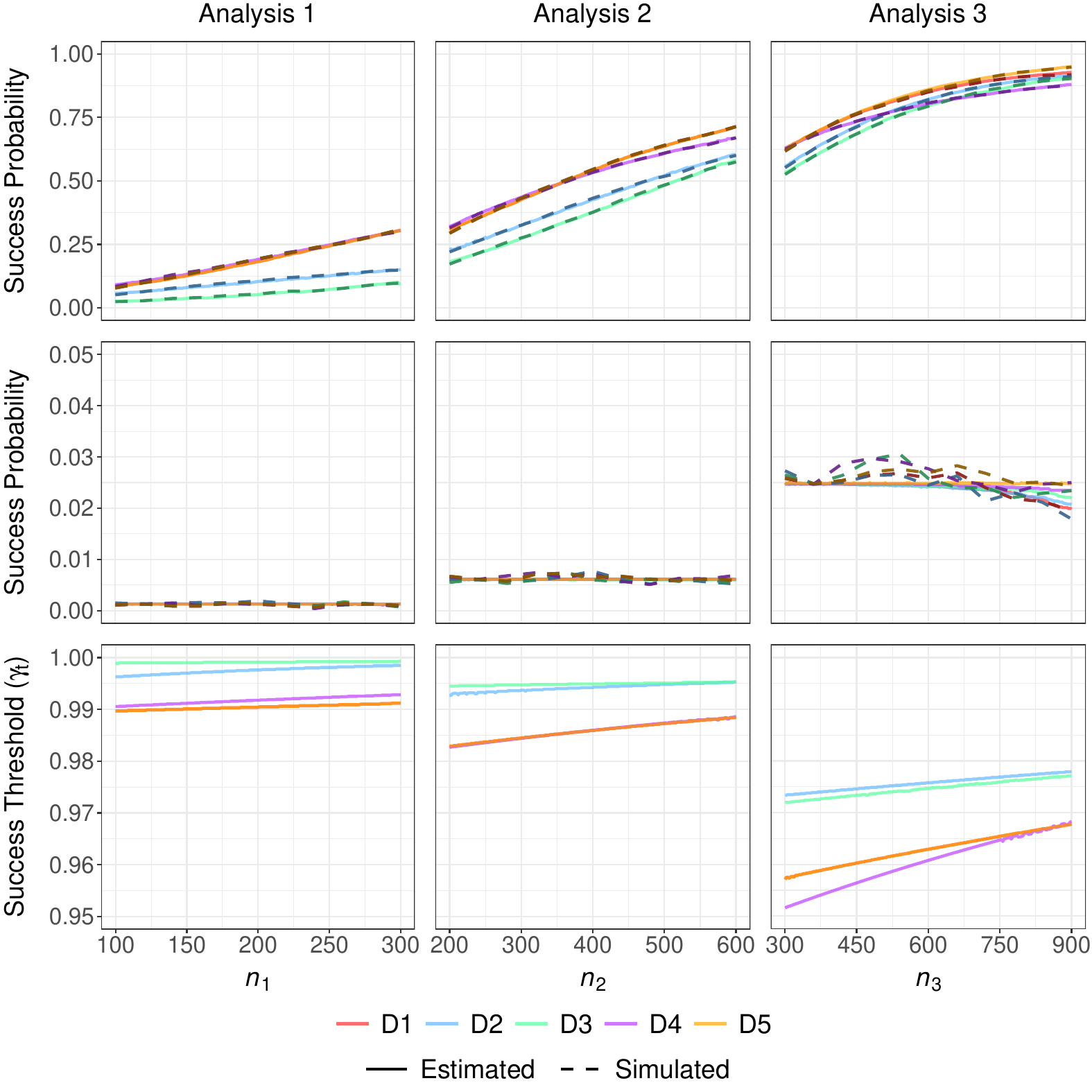} 

		\caption{\label{fig:eff} The cumulative probabilities of stopping for success for each design. Top: Power under $\Psi_1$. Middle: The type I error rate under $\Psi_0$. Bottom: Success Thresholds.} 
	\end{figure}

       For all designs, Algorithm \ref{alg1} was implemented with $R = 10^4$ iterations and $M = 10^3$ repetitions to estimate posterior predictive probabilities. Sample sizes for the first analysis of $n_a = 120$ and $n_b = 280$ were selected  based on PLATINUM-CAN to explore designs with varying power. Figure \ref{fig:eff} visualizes the cumulative probability of stopping for success at each analysis with respect to $n$ for all designs given our choices for $\Psi_1$ (top row) and $\Psi_0$ (middle row). The bottom row depicts how the success thresholds calibrated to our alpha-spending function change with $n$. In Figure \ref{fig:eff}, the solid curves were \emph{estimated} using the linear approximations in Algorithm \ref{alg1} based on simulations at only $n_a$ and $n_b$. The dashed curves (in a darker shade of the color corresponding to each design) were \emph{simulated} by generating samples $\mathcal{D}_n$ to estimate sampling distributions of $\boldsymbol{\tau}_{\hspace*{-1.25pt}*}(\mathcal{D}_{n})$ at $n = \{100, 120, \dots, 300 \}$.

\begin{figure}[!tb] \centering 
		\includegraphics[width = 0.8\textwidth]{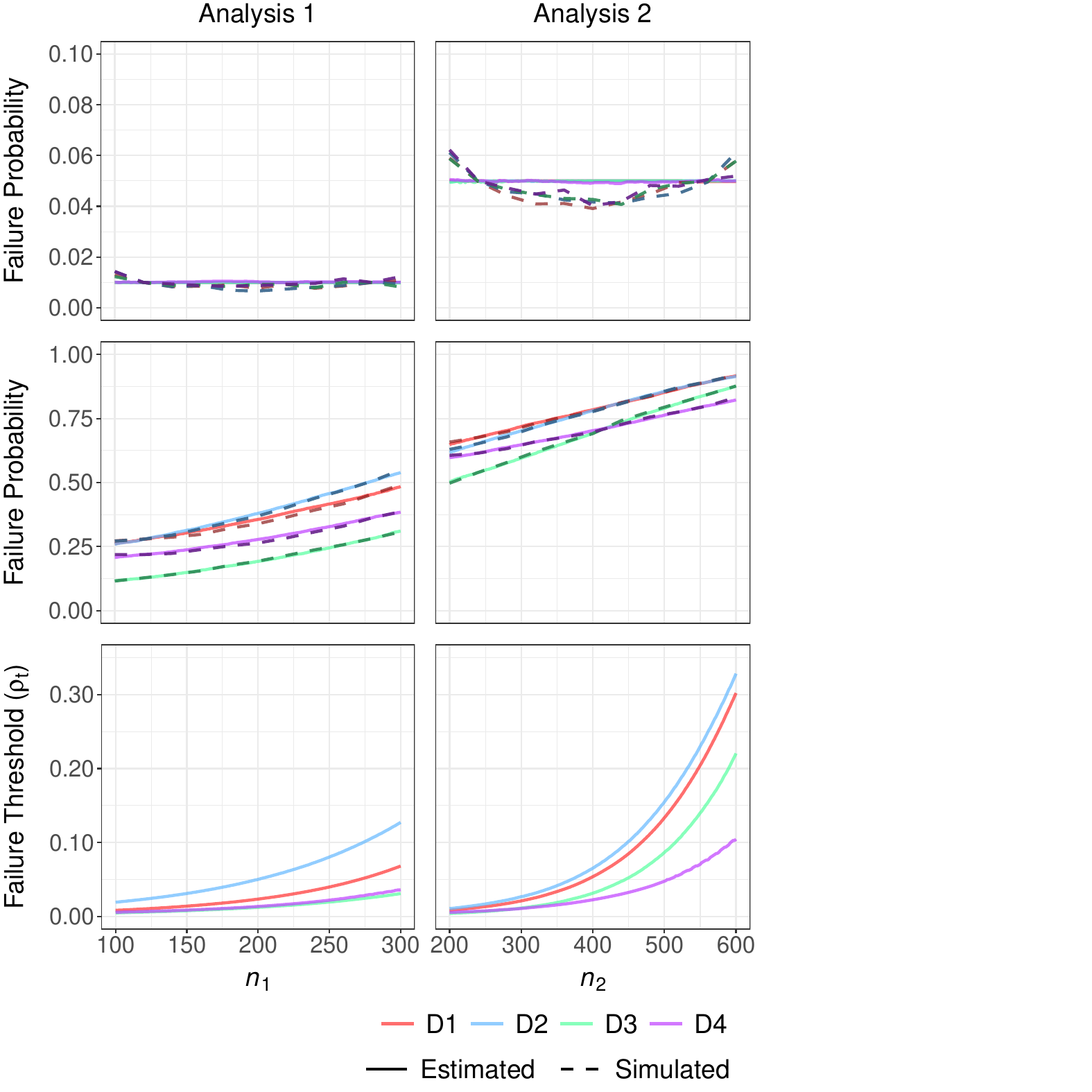} 

		\caption{\label{fig:fut} The cumulative probabilities of stopping for failure for each design. Top: Stopping under $\Psi_1$. Middle: Stopping under $\Psi_0$. Bottom: Failure Thresholds.} 
	\end{figure}

      Although the dashed curves are impacted by simulation variability, we use them as surrogates for the true stopping probabilities. We observe good alignment between the solid curves estimated using linear approximations and the dashed ones obtained by simulating new samples for this example. In the top row of Figure \ref{fig:eff}, the red curves for D1 and orange curves for D5 essentially coincide because there is a very low probability of stopping for failure at the first analysis (see Figure \ref{fig:fut}). In the middle row, the estimated probability of stopping for success is effectively constant at 0.0013 and 0.0061 for all sample sizes and designs, aligning with our alpha-spending function. At the final analysis, the estimated FWER is slightly less than the desired 0.025 for larger sample sizes in all designs except D5. As detailed in Section \ref{sec:power}, this behavior occurs because $\boldsymbol{\gamma}$ are tuned before accounting for stopping for failure.  Some estimated curves in Figure \ref{fig:eff} are slightly jagged because the decision thresholds are only being tuned to four decimal places, but these curves could be made smoother by calibrating the decision thresholds with greater precision. 

      In the bottom row of Figure \ref{fig:eff}, $\boldsymbol{\gamma}$ for D1 (red) and D5 (orange) coincide exactly. The success thresholds for D2 (blue) are more conservative than those for D1 and D5 due to the inflated type I error rate. The calibrated $\boldsymbol{\gamma}$ change as the sample size increases and the impact of the data overwhelms that of the informative prior in D1, D2, D4, and D5. The success thresholds also change slightly with $n$ for D3 with diffuse priors. Power in the top right subplot is the highest for D5 that does not allow early stopping for failure. Power is also higher for D1 than D2, where the observed data align better with the historical information. For D4, power is higher (lower) for smaller (larger) sample sizes, reflecting greater variability in the sampling distribution of $\boldsymbol{\tau}_{\hspace*{-1.25pt}*}(\mathcal{D}_{n})$ under the predictive approach.


     Figure \ref{fig:fut} visualizes the cumulative probability of stopping for failure at each analysis with respect to $n$ for D1 to D4 given our choices for $\Psi_1$ (top row) and $\Psi_0$ (middle row). The bottom row depicts the failure thresholds that are calibrated to our beta-spending function. Figure \ref{fig:fut} has no third column because the probability of stopping for failure at the final analysis can be directly obtained using the results in the final column of Figure \ref{fig:eff}. In the top row of Figure \ref{fig:fut}, the estimated probability of stopping for failure is effectively constant at 0.01 and 0.05 for all sample sizes and designs as enforced by binary search in Algorithm \ref{alg1}, aligning with the desired beta-spending function. The dashed simulated curves are somewhat U-shaped because the decision thresholds for all $n$ are tuned using linear approximations based on only the simulations at $n_a = 120$ and $n_b = 280$. These slight discrepancies between the solid and dashed curves occur because the quantiles of the true sampling distribution of $\boldsymbol{\tau}_{\hspace*{-1.25pt}*}(\mathcal{D}_{n})$ on the logit scale are not \emph{perfectly} linear functions of $n$, in part since the true posterior variances are not quite proportional to $n^{-1}$ for finite sample sizes due to the informative prior. In the middle row, the probability of failure increases with the sample size as the calibrated $\boldsymbol{\rho}$ become less conservative (bottom row). 
    
    The solid curves in Figures \ref{fig:eff} and \ref{fig:fut} are much easier to obtain since we need only estimate the sampling distribution of $\boldsymbol{\tau}_{\hspace*{-1.25pt}*}(\mathcal{D}_{n})$ at two values of $n$. 
    Even so, it took 1 hour on a high-computing server to estimate all solid curves for D1 to D4 in Figures \ref{fig:eff} and \ref{fig:fut} when approximating each posterior distribution using 5000 draws resulting from the conjugacy of each component of the mixture posterior. Results for D5 were obtained using the simulations for D1. We considered 11 values of $n$ to simulate the dashed curves for D1 to D4, taking 5.5 hours using the same computing resources. The computational savings scale with the number of sample sizes at which $\boldsymbol{\tau}_{\hspace*{-1.25pt}*}(\mathcal{D}_{n})$ is estimated. If a standard method estimates sampling distributions at $Q$ values of $n$, our approach is $Q/2$ times faster. Unlike standard methods, our approach allows practitioners to recalibrate decision thresholds and assess error probabilities for new values of $n$ without conducting additional simulations.

                \begin{table}[!b]
\centering
\begin{tabular}{ccccccc}
\hline
Design & $n$ & $\gamma_1$ & $\gamma_2$ & $\gamma_3$ & $\rho_1$ & $\rho_2$ \\ \hline
D1     & 200 & 0.9904     & 0.9859     & 0.9629     & 0.0234   & 0.0538   \\
D2     & 221 & 0.9978     & 0.9945     & 0.9762     & 0.0613   & 0.0948   \\
D3     & 239 & 0.9992     & 0.9951     & 0.9756     & 0.0176   & 0.0697   \\
D4     & 252 & 0.9924     & 0.9873     & 0.9644     & 0.0226   & 0.0500     \\
D5     & 194 & 0.9904     & 0.9859     & 0.9629     & ---      & ---      \\ \hline
\end{tabular}
\caption{Sample size $n$ for the first analysis, success thresholds $\boldsymbol{\gamma}$, and failure thresholds $\boldsymbol{\rho}$ recommended by Algorithm \ref{alg1} for each of the five designs.}
\label{tab:design}
\end{table}

    For all designs, the recommended decision thresholds and sample size $n$ for the first analysis are detailed in Table \ref{tab:design}. Given these recommendations for $n$, $\boldsymbol{\gamma}$, and $\boldsymbol{\rho}$, numerical results for all designs under $\Psi_0$ and $\Psi_1$ are summarized in Table \ref{tab:mpox}. These cumulative stopping probabilities were obtained using confirmatory simulations to estimate the joint sampling distribution of $\boldsymbol{\tau}_{\hspace*{-1.25pt}*}(\mathcal{D}_{n})$ at $\{n_t\}_{t=1}^T = n \times \{c_t\}_{t=1}^T$. The cumulative probabilities of stopping for success under $\Psi_0$ for all designs approximately coincide with the desired alpha-spending function of (0.0013, 0.0061, 0.0250), and power under $\Psi_1$ at the final analysis is about 0.85. Where applicable, the cumulative probabilities of stopping for failure are roughly 0.01 and 0.05 for the first two analyses, aligning with the selected beta-spending function. These results demonstrate the strong performance of our method.
    
        

\begin{table}[!htb]
\centering
\begin{tabular}{cccccccc}
\cline{3-7}
\multicolumn{1}{l}{} & \multicolumn{1}{l}{} & \multicolumn{5}{c}{Cumulative Stopping Probability}       & \multicolumn{1}{l}{} \\ \cline{1-8}
Design               & Model                & Success 1 & Success 2 & Success 3 & Failure 1 & Failure 2 & $\bar{n}$            \\ \hline
\multirow{2}{*}{D1}  & $\Psi_1$             & 0.1932    & 0.5452    & 0.8501    & 0.0077    & 0.0391    & 442.96               \\
                     & $\Psi_0$             & 0.0014    & 0.0065    & 0.0259    & 0.3390     & 0.7786    & 374.90               \\ \cline{2-8} 
\multirow{2}{*}{D2}  & $\Psi_1$             & 0.1179    & 0.4728    & 0.8538    & 0.0072    & 0.0382    & 472.78               \\
                     & $\Psi_0$             & 0.0010    & 0.0068    & 0.0247    & 0.4048    & 0.8072    & 356.04               \\ \cline{2-8} 
\multirow{2}{*}{D3}  & $\Psi_1$             & 0.0672    & 0.4603    & 0.8570    & 0.0073    & 0.0440    & 484.24               \\
                     & $\Psi_0$             & 0.0008    & 0.0040    & 0.0245    & 0.2346    & 0.7745    & 397.22               \\ \cline{2-8} 
\multirow{2}{*}{D4}  & $\Psi_1$             & 0.2578    & 0.6181    & 0.8538    & 0.0087    & 0.0441    & 414.26               \\
                     & $\Psi_0$             & 0.0010    & 0.0059    & 0.0259    & 0.3128    & 0.7622    & 383.62               \\ \cline{2-8} 
\multirow{2}{*}{D5}  & $\Psi_1$             & 0.1764    & 0.5424    & 0.8480    & ---       & ---       & 456.24               \\
                     & $\Psi_0$             & 0.0021    & 0.0074    & 0.0275    & ---       & ---       & 598.10               \\ \hline
\end{tabular}
\caption{Cumulative stopping probabilities for the recommended designs obtained by simulating confirmatory estimates of sampling distributions. The expected sample size ($\bar{n}$) is also provided.}
\label{tab:mpox}
\end{table}

\color{black}

    \subsection{Additional Examples}\label{sec:ex.2}

    \color{black}

   In Appendix D of the supplement, we consider design for the PLATINUM-CAN example using a time-to-event outcome without dynamic borrowing. In that example, we use Markov chain Monte Carlo methods for posterior approximation and illustrate how the sampling distribution estimates of $\boldsymbol{\tau}_{\hspace*{-1.25pt}*}(\mathcal{D}_{n})$ in Algorithm \ref{alg1} can be economically repurposed to consider multiple designs having a range of values for $T$. In Appendix E, we consider an industrial example that assesses the performance of our method in smaller sample settings. The strong performance of our method is confirmed in these additional examples.

\section{Discussion}\label{sec:disc}

 In this paper, we proposed an economical framework to estimate error probabilities associated with decision procedures based on posterior and posterior predictive probabilities in group sequential designs. This framework determines the minimum sample sizes that ensure power is sufficiently large \color{black} while calibrating the decision thresholds to selected alpha- or beta-spending functions. \color{black} The computational efficiency of our framework is predicated on a proxy for the joint sampling distribution of posterior summaries from a hypothetical GSD without early stopping criteria. We use the behavior in this large-sample proxy to motivate empirically estimating such sampling distributions at only two sample sizes. Our method significantly reduces the computational overhead required to design Bayesian group sequential experiments with early stopping rules in a wide range of disciplines.

 The methods proposed in this paper could be extended in various aspects to accommodate more complex sequential designs. \color{black} Our work in this article emphasized GSDs with at most two interventions. While the theory pertaining to our proxy sampling distributions can be applied to designs with more than two interventions, it is more complicated to define decision rules and error probabilities for such designs. \color{black} This paper also constrained the sample sizes for the analyses to be such that $\{n_t\}_{t=1}^T = n \times \{1, c_2, \dots, c_T\}$ as the sample size $n$ for the first analysis changes. \color{black} Our current contribution also precludes sequential designs where the allocation probabilities $\{A_t\}_{t=1}^T$ are not pre-specified across stages, \color{black}including designs with response-adaptive randomization that is often incorporated into clinical trials and multi-armed bandit experiments.  Future research could consider efficient design procedures for these extensions. 

Moreover, the proxy sampling distributions introduced in this article rely on various large-sample regularity conditions. 
The conditions for the BvM theorem, for instance, may not be required when using semi-parametric Bayesian methods for inference based on robust loss functions \citep{bissiri2016general} instead of standard Bayesian inference that relies on the full specification of a likelihood function. We could further broaden the impact of our methods by relaxing some of these regularity conditions to accommodate experimental design more broadly in future work.

 \section*{Supplementary Material}
These materials include the proofs of Lemmas \ref{lem0}, \ref{lem1}, and \ref{lem2} as well as Theorem \ref{thm1}, along with content for the additional examples in Section \ref{sec:ex}. The code to conduct the numerical studies in the paper is available online: \url{https://github.com/lmhagar/SeqDesign}.

	\section*{Funding Acknowledgement}
 
 Luke Hagar acknowledges the support of a postdoctoral fellowship from the Natural Sciences and Engineering Research Council of Canada (NSERC). Shirin Golchi acknowledges support from NSERC, Canadian Institute for Statistical Sciences (CANSSI), and Fonds de recherche du Québec - Santé (FRQS).
	





\bibliographystyle{chicago}


\end{document}